\documentclass[12pt]{iopart}
\usepackage{iopams}  
\usepackage{bm}
\usepackage{breqn}

\def \bzeta {\boldsymbol \zeta_{\alpha}}

\begin{document}

\title[Fourier-Hermite decomposition of the collisional Vlasov-Maxwell system]{Fourier-Hermite decomposition of the collisional 
Vlasov-Maxwell system: implications for the velocity space cascade}

\author{O. Pezzi$^{1,2,3}$, F. Valentini$^3$, S. Servidio$^3$, E. Camporeale$^4$ and P. Veltri$^3$}
 
\address{$^1$Gran Sasso Science Institute, Viale F. Crispi 7, I-67100 L'Aquila, Italy \\
$^2$INFN/Laboratori Nazionali del Gran Sasso, Via G. Acitelli 22, I-67100 Assergi (AQ), Italy \\
$^3$Dipartimento di Fisica, Universit\`a della Calabria, Via P. Bucci, I-87036 Arcavacata di Rende (CS), Italy\\
$^4$Center for Mathematics and Computer Science (CWI), 1090 GB Amsterdam, The Netherlands}

\ead{oreste.pezzi@gssi.it}

\vspace{10pt}
\begin{indented}
\item[]\today
\end{indented}

\begin{abstract}
Turbulence at kinetic scales is an unresolved and ubiquitous phenomenon that characterizes both space and laboratory plasmas. Recently, new 
theories, {\it in-situ} spacecraft observations and numerical simulations suggest a novel scenario for turbulence, characterized by a 
so-called phase space cascade -- the formation of fine structures, both in physical and velocity space. This new concept is here extended by 
directly taking into account the role of inter-particle collisions, modeled through the nonlinear Landau operator or the simplified 
Dougherty operator. The characteristic times, associated with inter-particle correlations, are derived in the above cases. The implications 
of introducing collisions on the phase space cascade are finally discussed.
\end{abstract}

%
\vspace{2pc}
%
\submitto{\PPCF}
%
%
%

\section{Introduction}
\label{sect:intro}
Hot and dilute plasmas, which are ubiquitous in near-Earth environments and in astrophysical systems, often exhibit a strongly 
turbulent and complex dynamics \cite{marsch97,bruno16,servidio15}. The fluctuations energy, injected at large scales, is typically 
transferred to smaller scales, enabling the energy transfer to particles. This process ultimately ceases by dissipating the energy 
and heating the plasma \cite{vaivads16}. When the energy of the fluctuations is available to be transferred to the distribution function 
of particles, it is possible to recover dynamical states which are very far from the thermal equilibrium. As a consequence, the particles 
distribution function is strongly deformed and exhibits temperature anisotropies, rings and beam-like structures as well as velocity 
space jets and vortices \cite{marsch06,matteini13,lesur14,haynes14,chen16,sorriso18a, sorriso18b}. Even though these systems are usually 
described by means of collisionless models, the presence of fine structures (i.e. strong gradients) in velocity space may enhance the role 
of collisions \cite{pezzi16a,pezzi17a}. Collisions may therefore result as one of the possible ingredients contributing to plasma heating 
\cite{pezzi16a,banonnavarro16,li16}. It should be highlighted that collisions have the intrinsic characteristic of making the system 
irreversible. Hence, it is possible - when considering collisional effects - to heat the system in a pure thermodynamic sense, 
intimately related to the irreversible degradation of the information.
 
Extensive efforts have been devoted to understand whether and how the turbulent cascade, routinely depicted in physical space, evolves also 
in velocity space. To better appreciate the presence of phase-space fluctuations, the particle distribution function is often decomposed in 
terms of Hermite polynomials for the velocity space variables, while the usual Fourier decomposition is adopted to describe fluctuations in 
physical space. Since the seminal work of Grad \cite{grad49}, numerous results have invoked the Hermite decomposition of the 
velocity distribution function. Several analyses have been focused on the description of the Vlasov-Poisson system in the collisionless 
case \cite{grant67,siminos11} or by modeling collisions with the Lenard-Bernstein operator 
\cite{lenard58,ng99,ng04,black13,plunk14,parker15,kanekar15,adkins18}. This system has been also studied extensively from a numerical 
viewpoint \cite{valentini05,camporeale06,valentini12,pezzi14,camporeale16,pezzi16b,vencels16}. Other important studies have been 
dedicated to the description of phase-space 
fluctuations in the framework of gyrokinetics and drift-wave turbulence 
\cite{hammett93,schekochihin08,tatsuno09,zocco11,teaca12,hatch13,parker16,schekochihin16} as well as, for fusion research, ion temperature 
gradient (ITG) driven turbulence \cite{watanabe02,watanabe04}. Even for this latter category, collisions are usually modeled through 
simplified collisional operators, such as the Lenard-Bernstein model. Very recently, an extensive characterization of the phase-space 
cascade has been proposed by Eyink \cite{eyink18}.

Separate studies have instead focused to model collisional effects by means of collisional integrals, such as the Landau one 
\cite{landau36}, which can be derived from ``first-principle'' arguments. In particular, the Hermite moments of the Landau integral and its 
gyrokinetics formulation have been recently discussed \cite{hirvijoki16,hirvijoki17,pfefferle17}. The analysis has been also extended to the 
hybrid or full Vlasov-Maxwell system. This has been done for numerical objectives \cite{delzanno15} and also for describing space plasmas 
\cite{servidio17,pezzi18,cerri18}. Indeed, complete models are crucial to describe kinetic turbulence in space plasmas, where complex 
phenomena - such as, for example, magnetic reconnection, stationary (zero-frequency) current structures and nonlinear damping - emerge at 
different spatial and temporal scales and interact 
\cite{perrone13,matthaeus14,servidio14,valentini16,valentini17,cerri17,groselj17,pezzi17b,pucci17,perrone18,parashar18,franci18,
hellinger18,olshevsky18,pucci18}.

The Hermite decomposition of the particle distribution function has been adopted to characterize the velocity space cascade. This picture 
indeed resembles a process of cascade where free energy is injected at large scales (low Hermite coefficients) and is transferred to large 
Hermite modes, analogously to the Fourier counterpart for the physical space cascade of fluids. At small velocity scales, finally, 
collisions ultimately provide the channel for dissipating these fluctuations. For the first time, in Servidio {\it et al.} 
\cite{servidio17}, thanks to the high-resolution measurements of Magnetospheric Multiscale Mission (MMS) \cite{burch16}, the velocity space 
cascade has been directly observed in the Earth's magnetosheath. A novel collisionless theory, based on the Kolmogorov approach, has been 
also developed and theoretical predictions are in accordance with both {\it in-situ} observations \cite{servidio17} and numerical results 
obtained within the hybrid Vlasov-Maxwell framework \cite{pezzi18,cerri18}.

Here we extend the theory proposed in Ref. \cite{servidio17} by taking into account the role of inter-particle collisions. Collisions are 
here modeled through the Landau operator \cite{landau36,rosenbluth57}, which can be derived from the Liouville equation, or, alternatively, 
the Dougherty operator \cite{dougherty64,dougherty67a,dougherty67b}, which is an ``ad-hoc'' simpler operator, which is still nonlinear and 
obeys the H-theorem. The Dougherty operator has been recently compared to the Landau one \cite{pezzi15a} and adopted for performing 
self-consistent Eulerian simulations \cite{pezzi15b,pezzi18b}. The two operators are here written in the Fourier-Hermite space and the 
differences between them are discussed in detail. It is shown that - in the asymptotic regime - the Landau operator shows two characteristic 
times, respectively associated with the diffusive and the drag part of the operator. The fastest characteristic time related to the Landau 
operator scales as $m^{-2}$, being $m$ the Hermite coefficient. On the other hand, the collisional characteristic time, when 
collisions are modeled through the Dougherty operator, is proportional to $m^{-1}$. For the Landau operator case, we derive the 
typical Hermite coefficients $m^*$, corresponding to the balance of the collisional characteristic time and the collisionless ones. 
We apply these results to typical natural and laboratory plasmas, finding that $m^*$ is generally large. This implies that the 
phenomenological theory described by Servidio {\it et al.} \cite{servidio17} holds its validity up to large Hermite coefficients.

As far as we know, the Fourier-Hermite decomposition of the Landau operator represents a novel result: previous works have been focused on 
the Hermite moments of the Landau operator  \cite{hirvijoki16,hirvijoki17,pfefferle17}, while here the relevant part of the discussion 
concerns the implication of including collisions in the phase space cascade. Moreover, here we present a compact notation of the collisional 
Vlasov-Maxwell system in terms of annihilation and creation operators, widely adopted in the quantum mechanics framework.

The paper is organized as follows. In Sect. \ref{sect:dec}, we revisit the Fourier-Hermite decomposition of the Vlasov-Maxwell system of 
equations, by ignoring collisions. Then, in Sect. \ref{sect:coll}, we include the effect of collisions, modeled through both the Landau 
operator or, alternatively, the Dougherty operator. In Sect. \ref{sect:impl} we discuss the role of collisions in terms of velocity space 
cascade by deriving the collisional characteristic time and by comparing it to the characteristic times associated with the 
collisionless part of the Vlasov equation. Finally, in Sect. \ref{sect:concl}, a summary of the presented results is given.

\section{Fourier-Hermite decomposition of Vlasov-Maxwell equations}
\label{sect:dec}
The dynamics of a non-relativistic, quasi-neutral plasma in absence of inter-particle collisions can be described by means of the 
Vlasov-Maxwell system of equations, which in CGS units are:
\begin{eqnarray}
\frac{\partial f_{\alpha}}{\partial t } + {\bm v} \cdot {\boldsymbol \nabla} f_{\alpha} + \frac{q_{\alpha}}{m_{\alpha}} \left( 
{\bm E} + \frac{{\bm v}}{c} \times {\bm B} \right) \cdot {\boldsymbol \nabla_{\bm v}} f_{\alpha} = 0 \label{eq:vlasov}\\
{\boldsymbol \nabla} \cdot {\bm E} = 4 \pi \rho \label{eq:divE} \\
{\boldsymbol \nabla} \cdot {\bm B} = 0 \label{eq:divB} \\
{\boldsymbol \nabla} \times {\bm E} = - \frac{1}{c} \frac{\partial {\bm B}}{\partial t} \label{eq:rotE} \\
{\boldsymbol \nabla} \times {\bm B} = \frac{1}{c} \frac{\partial {\bm E}}{\partial t} + \frac{4\pi}{c} {\bm j} \label{eq:rotB}
\end{eqnarray}
where $f_{\alpha}({\bm r}, {\bm v},t)$ is the distribution function of the $\alpha$-species, ${\bm E}({\bm r},t)$ and ${\bm B}({\bm 
r},t)$ are respectively the electric and magnetic fields, $q_\alpha$ and $m_\alpha$ are the charge and mass of the $\alpha$-species and
$\rho ({\bm r},t) = \sum_{\alpha} q_{\alpha} n_{\alpha} ({\bm r},t)$ and ${\bm j} ({\bm r},t)= \sum_{\alpha} q_{\alpha} n_{\alpha}({\bm 
r},t) {\bm u}_{\alpha}({\bm r},t)$ are the charge and current density, respectively. In the above definitions, $n_{\alpha}({\bm r},t)=\int 
d^3v f_{\alpha}({\bm r},{\bm v}, t)$ and ${\bm u}_{\alpha}({\bm r},t) = \int d^3v \, {\bm v} f_{\alpha}({\bm 
r},{\bm v}, t)/n_{\alpha}({\bm r},t)$ are respectively the $\alpha$-species number density and bulk speed. For the sake of simplicity, the 
dependencies on ${\bm r}$, ${\bm v}$ and $t$ of the several variables in Eq. (\ref{eq:vlasov})--(\ref{eq:rotB}) are omitted. Note that, at 
the right hand side of Eq. (\ref{eq:vlasov}), no collisional operators are introduced. 

In order to decompose Eqs. (\ref{eq:vlasov})--(\ref{eq:rotB}), we adopt the asymmetrically weighted Hermite functions, whose 
three-dimensional orthonormal basis is:
\begin{equation}
 \left\{ \eqalign{{\boldsymbol \Psi}_{\bm m} (\bzeta) &= \psi_{m_x}(\zeta_{\alpha,x})\, \psi_{m_y}(\zeta_{\alpha,y})\, 
\psi_{m_z}(\zeta_{\alpha,z}) \\
 {\boldsymbol \Psi}^{\bm m} (\bzeta) &= \psi^{m_x}(\zeta_{\alpha,x})\, \psi^{m_y}(\zeta_{\alpha,y})\, 
\psi^{m_z}(\zeta_{\alpha,z}) } \right. \label{eq:Psi}
\end{equation}
where $\bzeta= {\bm v} - {\bm u}_{\alpha,0}/\sqrt{2} v_{th\alpha,0}$, while ${\bm u}_{\alpha,0}$ and $v_{th\alpha,0}=\sqrt{k_B 
T_{\alpha,0}/m_\alpha}$ are the $\alpha$-species bulk and thermal speed at $t=0$. We also assume that (a) each species is initially at 
rest (${\bm u}_{\alpha,0}=0$) and (b) the initial equilibrium is homogeneous ($T_{\alpha,0}$ and $n_{\alpha,0}$ are constant). Clearly, the 
variable $\bzeta$ is independent from  ${\bm r}$ and $t$. 

In Eqs. (\ref{eq:Psi}), $\psi_m$ and $\psi^m$ are respectively the covariant and controvariant one-dimensional Hermite functions: 
\begin{equation}
\left\{ \eqalign{ \psi_{m} (\zeta) &= \frac{H_m(\zeta) e^{-\zeta^2}}{\sqrt{2^m m! \pi}} \\
 \psi^{m} (\zeta) &= \frac{H_m(\zeta)}{\sqrt{2^m m!}} } \right. \label{eq:psi}
\end{equation}
being $H_m(\zeta) = (-1)^m e^{\zeta^2} d^m/d\zeta^m e^{-\zeta^2}$ the $m$-th order ``physicists'' Hermite polynomial. The following 
properties are satisfied for the Hermite functions defined in Eqs. (\ref{eq:psi}):
\begin{equation}\label{eq:psipr}
\left\{ \eqalign{  &\int d\zeta \psi^m (\zeta) \psi_{n} (\zeta) = \delta_{m,n}  \\
 &\zeta \psi_m (\zeta) = \sqrt{\frac{m+1}{2}} \psi_{m+1}(\zeta) + \sqrt{\frac{m}{2}} \psi_{m-1} (\zeta)  \\ 
 &\frac{d \psi_m(\zeta)}{d \zeta} = - \sqrt{2(m+1)} \psi_{m+1} (\zeta) } \right.
\end{equation}

The above defined basis is exploited to decompose the variables in Eqs. (\ref{eq:vlasov})--(\ref{eq:rotB}) as follows:
\begin{equation}\label{eq:dec}
\left\{
\eqalign{ f_{\alpha} ({\bm r},{\bm v}, t) = \sum_{{\bm m},{\bm k}} \tilde{f}_{\alpha,{\bm m},{\bm k}}(t)\, e^{i {\bm k} \cdot {\bm r}}  
{\boldsymbol \Psi}_{\bm m} (\bzeta) \label{eq:decf} \\
g ({\bm r}, t) = \sum_{{\bm k}} {\tilde{g}_{{\bm k}}(t)\, e^{i {\bm k} \cdot {\bm r}} } } \right.
\end{equation}
being $g$ a generic function which depends only on ${\bm r}$ and $t$, such as $\rho$ or any component of ${\bm E}$, ${\bm B}$ or ${\bm j}$.

To obtain the evolution equation for the Fourier-Hermite coefficient $\tilde{f}_{\alpha,{\bm m},{\bm k}}$, Eq. (\ref{eq:vlasov}) is 
multiplied for ${\boldsymbol \Psi}^{\bm m}(\bzeta) e^{-i {\bm k} \cdot {\bm r}}\,d^3rd^3\zeta_\alpha$ and, then, the integral on 
the whole phase space is evaluated. After some algebra, it is easy to obtain:
\begin{equation}\label{eq:vlasovnk}
\eqalign{
\fl \frac{\partial {\tilde f}_{\alpha,{\bm m},{\bm k}}}{\partial t} + i v_{th\alpha,0}\, {\bm k} \cdot \left(\hat{\bm a}^{\dag} + \hat{\bm 
a} \right)  \tilde{f}_{\alpha,{\bm m},{\bm k}} -\frac{q_\alpha}{m_\alpha v_{th\alpha,0}}  \times  \\
\times \sum_{{\bm k}_2} \left( {\tilde {\bm E}}_{{\bm k}-{\bm k}_2} + \frac{v_{th\alpha,0}}{c} \hat{\bm 
a}^{\dag} \times {\tilde {\bm B}}_{{\bm k}-{\bm k}_2} \right) \cdot \hat{\bm a}  \tilde{f}_{\alpha,{\bm m},{\bm k}_2} = 0 }
\end{equation}
Since in Eq. (\ref{eq:vlasov}) the Lorentz-force term is nonlinear in space, the convolution over the Fourier wavevector ${\bm 
k_2}$ is recovered in Eq. (\ref{eq:vlasovnk}). It is worth to highlight that, with respect to the notation adopted in previous papers 
\cite{camporeale06,delzanno15}, here - in order to achieve a compact notation - we make use of the vectorial creation $\hat{\bm 
a}^{\dag}$ and annihilation $\hat{\bm a}$ operators, defined as usual as:
\begin{equation}
\left\{ \eqalign{\hat{ a}_j \tilde{f}_{\alpha,{\bm m},{\bm k}} = \sqrt{m_j}\, \tilde{f}_{\alpha,{\bm m} -{\bm e}_j,{\bm k}} 
\\
 \hat{ a}_j^\dag \tilde{f}_{\alpha,{\bm m},{\bm k}} = \sqrt{m_j + 1}\, \tilde{f}_{\alpha,{\bm m}+{\bm e}_j,{\bm k}}} 
\right.
\label{eq:crann} 
\end{equation}
being $j=x,y,z$ and ${\bm e}_i$ the $i$-th Cartesian unit vector. Each time a creation $\hat{\bm a}^{\dag}$ or annihilation $\hat{\bm a}$ 
operator is introduced, a factor proportional to $\sqrt{m}$ is recovered. Note also that in Eq. (\ref{eq:vlasovnk}) only neighbor couplings 
of different Hermite modes are recovered.

\section{Collisional operators in the Fourier-Hermite space}
\label{sect:coll}

In the current section we extend the result of Eq. (\ref{eq:vlasovnk}) by considering inter-particle collisions and, hence, evaluating
\begin{equation}\label{eq:Cank}
 \tilde{C}_{\alpha,{\bm m},{\bm k}}(t)=\int d^3\zeta_\alpha d^3r {\boldsymbol \Psi}^{\bm m}(\bzeta) e^{-i {\bm k} \cdot {\bm 
r}} C_\alpha
\end{equation}
where $C_\alpha$ is the collisional operator. 

\subsection{Collisions modeled with the Landau operator}

When focusing on the Landau operator, it is convenient to use the Landau operator written in terms of the Rosenbluth-MacDonald-Judd 
(RMJ) potentials \cite{rosenbluth57}:
\begin{equation}\label{eq:LANRMJ}
\eqalign{
 C^{LAN}_{\alpha} = \sum_{\beta} \frac{2 \pi q_\alpha^2 q_\beta^2 \ln \Lambda}{m_\alpha} \frac{\partial}{\partial v_i} \times \\
\times \left[ \frac{1}{2m_\alpha} \frac{\partial }{\partial v_j} \left(\frac{\partial^2 g_\beta}{\partial v_i \partial v_j} f_\alpha 
\right)- \frac{1}{\mu_{\alpha\beta}} \frac{\partial h_\beta}{\partial v_i} f_\alpha \right]  } 
\end{equation}
where
\begin{equation}\label{eq:RMJpot}
\left\{ \eqalign{g_\beta ({\bm r},{\bm v}, t) = \int d^3v' f_\beta({\bm r},{\bm v}',t)\, |{\bm v} - {\bm v}'|  \\
h_\beta ({\bm r},{\bm v}, t) = \int d^3v' \frac{f_\beta({\bm r},{\bm v}',t)}{|{\bm v} - {\bm v}'|} } \right.
\end{equation}
are the RMJ potentials and $\mu_{\alpha\beta}=m_\alpha m_\beta/(m_\alpha+m_\beta)$ is the reduced mass. This formulation allows to 
appreciate the Fokker-Planck structure of the Landau collisional operator.

By inserting Eq. (\ref{eq:LANRMJ}) in Eq. (\ref{eq:Cank}), one gets:
\begin{equation} \label{eq:CankLAN}
\eqalign {  \tilde{C}_{\alpha,{\bm m},{\bm k}}^{LAN}(t)= \sum_\beta \frac{2 \pi q_\alpha^2 q_\beta^2 \ln \Lambda}{m_\alpha 
v_{th\alpha,0}^2} \sum_{{\bm m}_1,{\bm m}_2,{\bm k}_2} \tilde{f}_{\alpha,{\bm m}_2,{\bm k}_2} \times  \\
\times \left[ \frac{ (\hat{ a}_i\hat{ a}_j g_{\beta,{\bm m}_1,{\bm k} - {\bm k}_2}) I_{{\bm m},{\bm m}_1,{\bm m}_2}}{2m_\alpha 
v_{th\alpha,0}^2} + \right. \\
\left. - \frac{ (\hat{ a}_i h_{\beta,{\bm m}_1,{\bm k} - {\bm k}_2}) J_{{\bm m},{\bm m}_1,{\bm m}_2}}{\mu_{\alpha\beta}} \right] }
\end{equation}
where:
\begingroup
\small
\begin{eqnarray}
&I_{{\bm m},{\bm m}_1,{\bm m}_2} = \int d^3\zeta_\alpha {\boldsymbol \Psi}^{\bm m} \left[ (\hat{ a}^\dag_i\hat{ a}^\dag_j {\boldsymbol 
\Psi}_{{\bm m}_1}) {\boldsymbol \Psi}_{{\bm m}_2} + \right.& \nonumber \\
&\hspace{1cm} \left. + (\hat{ a}^\dag_i {\boldsymbol \Psi}_{{\bm m}_1})(\hat{ a}^\dag_j {\boldsymbol \Psi}_{{\bm m}_2}) +  (\hat{ a}^\dag_j 
{\boldsymbol \Psi}_{{\bm m}_1})(\hat{ a}^\dag_i {\boldsymbol \Psi}_{{\bm m}_2}) + \right.& \label{eq:I} \\
&\hspace{4cm} \left. + {\boldsymbol \Psi}_{{\bm m}_1} (\hat{ a}^\dag_i\hat{ a}^\dag_j {\boldsymbol \Psi}_{{\bm m}_2}) \right]& \nonumber \\
&J_{{\bm m},{\bm m}_1,{\bm m}_2} = \int d^3\zeta_\alpha {\boldsymbol \Psi}^{\bm m} \left[ (\hat{ a}^\dag_i {\boldsymbol 
\Psi}_{{\bm m}_1}) {\boldsymbol \Psi}_{{\bm m}_2} + \right. \nonumber \\
&\hspace{4cm} \left. + {\boldsymbol \Psi}_{{\bm m}_1} (\hat{ a}^\dag_i {\boldsymbol \Psi}_{{\bm m}_2}) 
\right] \label{eq:J}
\end{eqnarray}
\endgroup
The last two integrals contain the product of three Hermite functions and they are non-null in the case of even summation of the 
three involved Hermite coefficients \cite{gradshteyn}. In the asymptotic regime of large $m$ ($m\sim m\pm 1$), the two integrals have the 
following dependence on the Hermite coefficient $m$: $I_{{\bm m},{\bm m}_1,{\bm m}_2} \sim 
m$ and $J_{{\bm m},{\bm m}_1,{\bm m}_2} \sim \sqrt{m}$. 

By looking at Eq. (\ref{eq:CankLAN}), one easily realizes that nonlinearities of the Landau operator (i.e. the form of its Fokker-Planck 
coefficients) explicitly depend on the velocity coordinates. Therefore, Eq. (\ref{eq:CankLAN}) exhibits the convolutions over the Hermite 
coefficients. The Landau operator is hence non-local in the Hermite space: for a given Hermite coefficient ${\bm m}$, the Landau 
operator affects also the other Hermite modes. This represents a peculiar characteristic of the Landau operator, which is lost in other 
simplified operators, such as the Lenard-Bernstein or the Dougherty operator. In other words, the local coupling of Hermite modes, 
typical of collisionless systems [Eq. (\ref{eq:vlasovnk})], is drastically modified into a global coupling when introducing the Landau 
operator.

\subsection{Collisions modeled through the Dougherty operator}
Here we shift our focus on the case of the Dougherty operator, whose expression is:
\begin{equation}\label{eq:DG}
\eqalign{ C^{DG}_{\alpha} ({\bm r},{\bm v}, t) = \sum_{\beta} \nu_{\alpha\beta}({\bm r},t) \frac{\partial}{\partial v_i} \left[\frac{k_B 
T_{\alpha\beta}({\bm r},t)}{m_\alpha} \times \right. \nonumber \\
\left. \times \frac{\partial f_\alpha({\bm r},{\bm v}, t) }{\partial v_i} + \left(v_i - u_{\alpha\beta,i}({\bm r},t)\right) f_{\alpha}({\bm 
r},{\bm v}, t) \right] } 
\end{equation}
The collisional frequency $\nu_{\alpha\beta}({\bm r},t)$, the generalized speed ${\bm u}_{\alpha\beta}({\bm r},t)$ and 
the generalized temperature $T_{\alpha\beta}({\bm r},t)$ in Eq. (\ref{eq:DG}) are obtained by adopting a ``simple'' Fokker-Planck structure 
for the Dougherty operator and by expanding the Landau operator around an equilibrium distribution function $f_0$. Finally, by comparing the 
energy and momentum transfer equations obtained for the Landau and the Dougherty operators, it is possible to set the proper values for the 
parameters. (See Ref. \cite{dougherty67a} for a more detailed discussion). 

We notice that the Dougherty operator nonlinearities are intrinsically different from the Landau operator ones, since the Fokker-Planck 
coefficients of the Dougherty operator do not explicitly depend on the velocity coordinates. Therefore, contrary to the Landau operator 
case, we do not here expect to recover convolutions in the Hermite space. Indeed, by decomposing the Dougherty operator, one gets:
\begingroup\small
\begin{equation} \label{eq:CankDG}
\eqalign{ \tilde{C}_{\alpha,{\bm m},{\bm k}}^{DG}(t)= \sum_\beta \sum_{{\bm k}_2, {\bm k_3}} \tilde{\nu}_{\alpha\beta,{\bm 
k}-{\bm k}_2-{\bm k}_3} \times  \\ 
\times \left[ \left( \frac{{\tilde T}_{\alpha\beta,{\bm k}_3}}{T_{\alpha,0}} - \delta({\bm k}_3) \right) \left(\hat{\bm a} \cdot 
\hat{\bm a}\right) - \left( {\tilde {\bm u}}_{\alpha\beta,{\bm k_3}} \cdot \hat{\bm a}\right) \right] \tilde{f}_{\alpha,{\bm m},{\bm 
k}_2}+\\
- \sum_\beta \sum_{{\bm k}_2} \tilde{\nu}_{\alpha\beta,{\bm k}-{\bm k}_2} \left( \hat{\bm a}^\dag \cdot \hat{\bm a}\right) 
\tilde{f}_{\alpha,{\bm m},{\bm k}_2} 
}
\end{equation}
\endgroup
where $\tilde{\nu}_{\alpha\beta,{\bm k}}$, ${\tilde T}_{\alpha\beta,{\bm k}}$ and $\tilde{\bm u}_{\alpha\beta,{\bm k}}$ are the Fourier 
coefficients of $\nu_{\alpha\beta}$, ${\bm u}_{\alpha\beta}$ and $T_{\alpha\beta}$ and the first term in Eq. (\ref{eq:CankDG}) includes the 
spatial convolution due to the spatial dependence of ${\bm u}_{\alpha\beta}$ and $T_{\alpha\beta}$. 

The behavior of the linear Lenard-Bernstein operator is recovered by linearizing the Dougherty operator. In this case, one has 
$T_{\alpha\beta}=T_{\alpha,0}$, ${\bm u}_{\alpha\beta} = {\bm u}_{\alpha 0}=0$ and $\nu_{\alpha\beta}=\nu_{\alpha\beta,0}$ and, hence 
${\tilde T}_{\alpha\beta,{\bm k}}=T_{\alpha,0}\, \delta({\bm k})$ and ${\tilde {\bm u}}_{\alpha\beta,{\bm k}}=0$ and ${\tilde 
\nu}_{\alpha\beta,{\bm k}}=\nu_{\alpha\beta,0}\,\delta({\bm k})$. Therefore, Eq. (\ref{eq:CankDG}) reduces to:
\begin{equation} \label{eq:CankDGlin}
\eqalign{
 \tilde{C}_{\alpha,{\bm m},{\bm k}}^{DG,lin}(t)= - \nu_{\alpha\beta,0} \left( \hat{\bm a}^\dag \cdot \hat{\bm a}\right) 
\tilde{f}_{\alpha,{\bm m},{\bm k}} = \\
 = - \nu_{\alpha\beta,0} \left( m_x + m_y + m_z \right) \tilde{f}_{\alpha,{\bm m},{\bm k}} }
\end{equation}
As the Lenard-Bernstein operator, the linearized Dougherty operator is diagonal in the Fourier-Hermite space, with eigenvalue $- 
\nu_{\alpha\beta,0} (m_x+m_y+m_z)$ \cite{ng99,ng04}. We again emphasize that, as expected, the Dougherty operator acts locally in the 
Hermite space.

\section{Implications for velocity space enstrophy cascade}
\label{sect:impl}
To analyze the effect of collisions in the Fourier-Hermite space, we here obtain the equation for the Hermite spectrum $E_{\alpha, {\bm 
m}, {\bm k}}=|\tilde{f}_{\alpha,{\bm m},{\bm k}}|^2$ \cite{parker16} (the definition of the Hermite spectrum here adopted is slightly 
different from the one of Ref. \cite{kanekar15,schekochihin16,servidio17}). By manipulating Eq. (\ref{eq:vlasovnk}) coupled with Eq. 
(\ref{eq:CankLAN}) or, alternatively, with Eq. (\ref{eq:CankDG}), the final result is:
\begin{equation}\label{eq:Emk}
 \frac{\partial E_{\alpha, {\bm m}, {\bm k}}}{\partial t} + \Lambda^{A,\alpha}_{{\bm m}, {\bm k}} - \Lambda^{E,\alpha}_{{\bm m}, 
{\bm k}} - \Lambda^{B,\alpha}_{{\bm m}, {\bm k}} = \Lambda^{\nu,\alpha}_{{\bm m}, {\bm k}} 
\end{equation}
where:
\begingroup\small
\begin{equation} \label{eq:lambda}
 \left\{ \eqalign{ 
 \Lambda^{A,\alpha}_{{\bm m}, {\bm k}} = i\, v_{th\alpha,0}\,  {\bm k} \cdot \left[ \tilde{f}^*_{\alpha,{\bm m},{\bm k}} 
\left(\hat{\bm a}^\dag+\hat{\bm 
a}\right)\tilde{f}_{\alpha,{\bm m},{\bm k}} \right] + c.c. \\
 \Lambda^{E,\alpha}_{{\bm m}, {\bm k}} = \frac{q_\alpha \tilde{f}^*_{\alpha,{\bm m},{\bm k}}}{m_\alpha v_{th\alpha,0}}  \sum_{{\bm 
k}_2} \left( \tilde{\bm E}_{{\bm k}-{\bm k}_2} 
\cdot \hat{\bm a} \right) \tilde{f}_{\alpha,{\bm m},{\bm k}} + c.c. \\
 \Lambda^{B,\alpha}_{{\bm m}, {\bm k}} =  \frac{q_\alpha \tilde{f}^*_{\alpha,{\bm m},{\bm k}}}{m_\alpha c}   \sum_{{\bm k}_2} 
\left( \hat{\bm a}^\dag \times \tilde{\bm 
B}_{{\bm k}-{\bm k}_2} \right) \cdot \hat{\bm a} \tilde{f}_{\alpha,{\bm m},{\bm k}} + c.c. \\
\Lambda^{\nu,\alpha}_{{\bm m}, {\bm k}} = \tilde{f}^*_{\alpha,{\bm m},{\bm k}} \tilde{C}_{\alpha,{\bm m},{\bm k}} + c.c.
}  \right.
\end{equation}
\endgroup
and c.c. indicates the complex conjugate and $\tilde{C}_{\alpha,{\bm m},{\bm k}}=\tilde{C}_{\alpha,{\bm m},{\bm k}}^{LAN}$ [Eq. 
(\ref{eq:CankLAN})] or $\tilde{C}_{\alpha,{\bm 
m},{\bm k}}=\tilde{C}_{\alpha,{\bm m},{\bm k}}^{DG}$ [Eq. (\ref{eq:CankDG})]. Note that, in the linearized Dougherty operator case, 
$\Lambda^{\nu,\alpha}_{{\bm m}, {\bm k}} = - \nu_{\alpha\beta,0} \left( m_x + m_y + m_z \right) E_{\alpha, {\bm m}, {\bm k}}$.

By manipulating the operators given in Eq. (\ref{eq:lambda}) as $\Lambda^{\alpha}_{{\bm m}, {\bm k}} \sim E_{\alpha, {\bm m}, {\bm 
k}} /\tau^{\alpha}_{m,k}$, one can obtain the characteristic times $\tau^{\alpha}_{m,k}$ associated with each process, in the asymptotic 
regime of large $m$. We assume that each direction in both physical and velocity spaces is equivalent: $m_x\simeq m_y \simeq m_z \simeq m$ 
and $k_x\simeq k_y \simeq k_z \simeq k$. Since we are interested in phenomenological scaling with $m$ and $k$, we neglect the presence of 
convolutions in both Fourier and Hermite spaces. We also specialize in the case of a quasi-neutral plasma composed by protons and 
electrons with similar temperature ($T_p\simeq T_e$) and we focus on the proton dynamics ($\alpha=p$), thus simplifying the collisional 
operator structure by taking into account proton-proton collisions. This case is of particular interest for solar wind and inter-stellar 
medium applications. Bearing in mind that a factor $\sqrt{m}$ occurs each time an annihilation or creation operator is present, we obtain:
\begin{equation}\label{eq:tau}
\left\{ \eqalign{   \tau^{A,p}_{m,k} \sim 1/v_{thp,0}k \sqrt{m}\\
\tau^{E,p}_{m,k} \sim m_p v_{thp,0}/e E_k \sqrt{m} \\
\tau^{B,p}_{m,k} \sim m_p c/e B_k m \\ 
\tau^{LAN,p}_{m,k} \sim \tau_{pp} (1/m^2 - 1/m) } \right.
\end{equation}
where $\tau^{A,p}_{m,k}$, $\tau^{E,p}_{m,k}$, $\tau^{B,p}_{m,k}$ and $\tau^{LAN,p}_{m,k}$ are respectively the characteristic times 
associated with advection, electric field, magnetic field and collision terms in Eq. (\ref{eq:Emk}) and $\tau_{pp} = m_p^2 v_{thp,0}^3/2\pi 
e^4 n_p ln\Lambda$ is the collisional time obtained by neglecting local velocity space effects, i.e. the ``global'' collisional time. Note 
that $m_p$ is the proton mass and has not be confused with the Hermite coefficient $m$. 

The collisionless characteristic times here derived differ from the ones obtained in Ref. \cite{servidio17}, since we do not average on the 
spatial domain and, hence, characteristic times also depend on the wavenumber $k$. The collisional characteristic time 
$\tau^{LAN,p}_{m,k}$ shows two different scaling with $m$: the diffusive (drag) term produces the scaling $m^{-2}$ ($m^{-1}$). At large 
$m$, the fastest contribution is due to the diffusive term:  $\tau^{LAN,p}_{m,k} \sim \tau_{pp}/m^2$. When focusing on the Dougherty 
operator case, in both linear and nonlinear regimes, the scaling is instead always $m^{-1}$.

It is possible to compare the characteristic times reported above to find Hermite coefficient $m^*$ (at a given wavenumber $k$) when the 
plasma dynamics changes from a collisionless regime to a collisional one. We remark that, the theory in Ref. \cite{servidio17} is based on 
the conservation of enstrophy (or free energy) $\Omega = \int d^3v \delta f^2$ and this assumption breaks when introducing collisions. 
Indeed, by considering both the Landau or the Dougherty operators - which hold the H-theorem for the entropy growth - the enstrophy is not 
anymore preserved. In other words, $m^*$ corresponds to the velocity space scale at which the phenomenological theory described by Servidio 
et al. \cite{servidio17} breaks its validity:
\begin{equation}\label{eq:m}
\left\{ \eqalign{   m_A^* \simeq (v_{thp,0}k \tau_{pp})^{2/3} \\
m_E^* \simeq (\tau_{pp} e E_k /m_p v_{thp,0})^{2/3} \\
m^*_B \simeq (\tau_{pp} e B_k /m_p c) } \right.
\end{equation}
where $m_A^*$, $m_E^*$ and $m_B^*$ are respectively associated to the advection, electric field and magnetic field terms in Eq. 
(\ref{eq:Emk}), i.e. are respectively significant if the advection, electric field or magnetic field terms are dominant in the left-hand 
side of Eq. (\ref{eq:Emk}). In general, $m^*$ decreases if the collisional characteristic time $\tau_{pp}$ decreases (i.e. collisions are 
faster), while it increases if the corresponding collisionless dominant term becomes more important (i.e. $E_k$, $B_k$ or $k$ larger). To 
appreciate the role of spatial fluctuations, we further simplify expressions in Eqs. (\ref{eq:m}) by assuming a fully-developed turbulent 
scenario with a Kolmogorov scaling for both velocity and magnetic field fluctuations: $u_k = u_c (k/k_c)^{-1/3}$ and $B_k = B_c 
(k/k_c)^{-1/3}$, being $u_c = u(k_c)$ and $B_c = B(k_c)$ the fluctuation amplitudes at the correlation scale of the turbulence 
$L_c=1/k_c$. The MHD scaling $E_k = u_k B_0/c$ is also assumed for the electric field fluctuations where $B_0$ is the background magnetic 
field. This choice is motivated by the fact that natural and laboratory plasmas are often strongly turbulent. Within these assumptions, it 
is easy to get:
\begin{equation}\label{eq:m2}
\left\{ \eqalign{   m_A^* \simeq (\sqrt{\beta_p} \; \Omega_{cp}\tau_{pp}\; k d_p )^{2/3} \\
m_E^* \simeq \left( \Omega_{cp}\tau_{pp}\, \frac{1}{\sqrt{\beta_p}}\; \frac{u_c}{c_A} \left(\frac{k}{k_c}\right)^{-1/3}\right)^{2/3} \\
m^*_B \simeq \left( \Omega_{cp}\tau_{pp}\, \frac{B_c}{B_0} \left(\frac{k}{k_c}\right)^{-1/3} \right) } \right.
\end{equation}
where $\Omega_{cp}= e B_0/m_p c$ is the proton cyclotron frequency, $d_p= c_A/\Omega_{cp}$ is the proton skin depth, $c_A=B_0/\sqrt{4\pi 
n_0 m_p}$ is the Alfv\'en speed, $\beta_p= 2 v_{thp}^2/c_A^2$ is the proton beta parameter and $n_0$ is the background density. The Hermite 
coefficients $m_A^*$, $m_E^*$ and $m_B^*$ depend on (i) the global collisional time normalized to the cyclotron time 
$\Omega_{cp}\tau_{pp}$, (ii) the proton beta $\beta_p$, (iii) the strength of turbulence $\eta \sim B_c/B_0 \sim u_c/c_A$ and (iv) the 
wavenumber $k$ at which fluctuations are evaluated. Both $m_E^*$ and $m_B^*$ decrease as $k\gg k_c$: as turbulence produces smaller scale 
fluctuations the characteristic Hermite coefficients, relevant for turning on collisions, gets shorter. This last aspect represents, from 
a different point of view, the collisionality enhancement due to the presence of fine velocity scale structures, mainly produced by 
turbulent fluctuations that perturb the particle distribution function.

\begin{table}[!t]
\caption{\label{tab:A} Physical parameters, necessary to evaluate $m_A^*$, $m_E^*$ and $m_B^*$, for the solar wind (first row), for 
the inter-stellar medium (center row) and for the hot-ion collisionless tokamak plasma (bottom row).}
\begin{indented}
\item[]\begin{tabular}{@{}lllllll}
\br
Plasmas&$n_0$&$B_0$&$T_p$&$L_c$&$\eta$&$k$\\
\mr
Slow solar wind			&$15 cm^{-3}$ 		 &$6 nT$	&$5\times 10^{4} K$	&$0.02 au$	&$1$	&$d_p^{-1}$\\
Inter-stellar medium		&$3\times10^{-3}cm^{-3}$ &$10^{-6}G$	&$10^6 K$		&$1 kpc$	&$0.5$	&$\rho_p^{-1}$\\
Hot-ion collisionless tokamak	&$4\times10^{13} cm^{-3}$&$2 T$		&$2\times 10^{3}keV$	&$0.25 m$	&$0.05$	&$\rho_p^{-1}$\\
\br
\end{tabular}
\end{indented}
\end{table}

We conclude the paper by calculating $m_A^*$, $m_E^*$ and $m_B^*$ for three typical weakly-collisional plasmas: the slow solar wind, the 
inter-stellar medium and the hot-ion collisionless tokamak plasma, which often shows a turbulent dynamics stirred by ion temperature 
gradient. Relevant parameters for the calculation are extrapolated by Refs. \cite{matthaeus82, sorriso99, matthaeus05, bruno16} for the 
solar wind; by Ref. \cite{lagage83} for the inter-stellar medium and by Refs. \cite{falchetto04,falchetto08} for the hot-ion collisionless 
tokamak plasma and are listed in Table \ref{tab:A}. In each case, the wavenumber $k$ is close to the proton inertial scales (being $\rho_p 
= v_{thp}/\Omega_{cp}$ the proton gyroradius).

Results are reported in Table \ref{tab:B}. In each system, the resulting Hermite coefficients useful to turn on collisions are quite large. 
This implies that: (i) the theory developed in Ref. \cite{servidio17} is valid up to these large Hermite coefficients; (ii) the presence of 
smaller scale spatial fluctuations may reduce $m^*$; (iii) potential failures of the collisionless assumption, induced by collisionality 
enhancement due to fine velocity space structures, occur whether the distribution function exhibits structures such that their associated 
Hermite coefficient is $m \sim200$. Recovering distribution function with such highly structured velocity space perturbations is not 
nowadays possible due to velocity space resolution limitations present in both spacecraft instruments and numerical simulations. 

\begin{table}[!htb]
\caption{\label{tab:B} Hermite coefficients $m_A^*$, $m_E^*$ and $m_B^*$ for the solar wind (first row), for the inter-stellar medium 
(center row) and for the hot-ion collisionless plasma (bottom row).}
\begin{indented}
\item[]\begin{tabular}{@{}llll}
\br
Plasmas&$m_A^*$&$m_E^*$&$m_B^*$\\
\mr
Slow solar wind&$1483$&$165$&$1800$\\
Inter-stellar medium&$10^6$&$325$&$19000$\\
Hot-ion collisionless tokamak&$384$&$471$&$960$\\
\br
\end{tabular}
\end{indented}
\end{table}

\section{Conclusions}
\label{sect:concl}
In this paper, the collisional Vlasov-Maxwell system of equations has been decomposed in the Fourier-Hermite space. This approach is 
extremely useful to describe fluctuations in velocity space and the coupling of turbulent fluctuations in both physical and velocity space. 
A compact notation, in terms of annihilation and creation operators, has been introduced. By modeling collisions through the Landau and the 
Dougherty (both nonlinear and linearized) operators, we have also decomposed the collisional operator in the Fourier-Hermite space. The 
features of the operators have been compared in detail. Finally we have shown that, by obtaining the equation for the Hermite spectrum 
$E_{\alpha, {\bm m}, {\bm k}}$, it is possible to derive the scaling of each term in the collisional Vlasov equation. The 
characteristic times associated with each part of the Vlasov-Landau equation have been derived. These times are local in the Fourier-Hermite 
space and they can give insights on how fluctuations in both physical and velocity spaces locally affect the relative importance of each 
term in the collisional Vlasov equation. The Hermite coefficients $m^*$, corresponding to the balance of the collisional time with the 
collisionless ones, have been derived separately for the advection, electric field and magnetic field terms. Under some assumptions, it has 
been possible to write simple expressions for $m^*$, that have been evaluated for three natural and laboratory plasmas: the slow solar 
wind, the inter-stellar medium and the hot-ion collisionless tokamak plasma. The resulting Hermite coefficients $m^*$, which correspond to 
the transition from a collisionless to a collisional regime in the plasma dynamics, are quite large for the considered systems. It is 
worth to note that the introduction of a collisional operator breaks the enstrophy conservation at small velocity scales, a principle which 
is analogous to the Kolmogorov cascade of energy for fluids.

\ack
O.P. and S.S. were partly supported by the International Space Science Institute (ISSI) in the framework of International Team 405 entitled 
“Current Sheets, Turbulence, Structures and Particle Acceleration in the Heliosphere”. E.C. was partially supported by NWO--Vidi grant 
639.072.716. O.P. sincerely thanks Prof. L. Sorriso-Valvo for the enjoyable and fruitful discussions on these topics. \\


\begin{thebibliography}{99}
\bibitem{marsch97} Marsch E and Tu CY 1997 {\em Nonlinear Processes in Geophysics} {\bf 4}, 101.
%
\bibitem{bruno16} Bruno R and Carbone V 2016 {\it Turbulence in the Solar Wind} vol. 928 (Berlin: Springer)
%
\bibitem{servidio15} Servidio S, Valentini F, Perrone D, Greco A, Califano F, Matthaeus WH and Veltri P 2015 {\em J. Plasma Phys.} 
{\bf 81} 328510107
%
\bibitem{vaivads16} Vaivads A, Retin\'o A, Soucek J, Khotyaintsev YuV, Valentini F, Escoubet CP, \etal 2016 {\em J. Plasma Phys.} 
{\bf 82} 905820501
%
\bibitem{marsch06} Marsch E 2006 {\em Living Reviews in Solar Physics} {\bf 3} 1
%
\bibitem{matteini13} Matteini L, Hellinger P, Goldstein BE, Landi S, Velli M and Neugebauer M 2013 {\em J. Geophys. Res.: Space Phys.} {\bf 118} 2771--2782
%
\bibitem{lesur14} Lesur M, Diamond PH and Kosuga Y 2014 {\em Phys. Plasmas} {\bf 21} 112307
%
\bibitem{haynes14} Haynes CT, Burgess D and Camporeale E 2014 {\em Astrophys. J.} {\bf 783}, 1, 38 
%
\bibitem{chen16} Chen CHK, Matteini L, Schekochihin AA, Stevens ML, Salem CL, Maruca BA, Kunz MW and Bale SD 2016 {\em Astrophys. J. Lett.} {\bf 825} L26 
%
\bibitem{sorriso18a} Sorriso-Valvo L, Perrone D, Pezzi O, Valentini F, Servidio S, Zouganelis Y and Veltri P 2018 {\em J. Plasma Phys.} {\bf 84} 725840201 
%
\bibitem{sorriso18b} Sorriso-Valvo L, Catapano F, Retin\`o A, Le Contel O, Perrone D, Roberts OW, ... and Khotyaintsev YuV 2018 {\it Turbulence-driven ion beams in the magnetospheric Kelvin-Helmholtz instability}, under review.
%
\bibitem{pezzi16a} Pezzi O, Valentini F and Veltri P 2016 {\em Phys. Rev. Lett.} {\bf 116} 145001
%
\bibitem{pezzi17a} Pezzi O 2017 {\em J. Plasma Phys.} {\bf 83}, 555830301
%
\bibitem{banonnavarro16} Navarro AB, Teaca B, Told D, Groselj D, Crandall P and Jenko F 2016 {\em Phys. Rev. Lett.} {\bf 117} 245101
%
\bibitem{li16} Li TC, Howes GG, Klein KG, TenBarge JM 2016 {\em Astrophys. J. Lett.} {\bf 832} L24

\bibitem{grad49} Grad H 1949 {\em Comm. pure and appl. math.} {\bf 2} 331--407
%
\bibitem{grant67} Grant FC and Feix MR 1967 {\em Phys. Fluids} {\bf 10} 696--702
%
\bibitem{siminos11} Siminos E, Benisti D and Gremillet L 2011 {\em Phys. Rev. E} {\bf 83.5} 056402
%
\bibitem{lenard58} Lenard A and Bernstein IB 1958 {\em Phys. Rev.} {\bf 112} 1456--1459
%
\bibitem{ng99} Ng CS, Bhattacharjee A and Skiff F 1999 {\em Phys. Rev. Lett.} {\bf 83.10} 1974 
%
\bibitem{ng04} Ng CS, Bhattacharjee A and Skiff F 2004 {\em Phys. Rev. Lett.} {\bf 92.6} 065002
%
\bibitem{black13} Black C, Germaschewksi K, Bhattacharjee A and Ng CS 2013 {\em Phys. Plasmas} {\bf 20.1} 012125
%
\bibitem{plunk14} Plunk GG and Parker JT 2014 {\em Eur. Phys. J D} {\bf 68} 296
%
\bibitem{parker15} Parker JT and Dellar PJ 2014 {\em J. Plasma Phys.} {\bf 81} 305810203
%
\bibitem{kanekar15} Kanekar A, Schekochihin AA, Dorland W and Loureiro NF 2015 {\em J. Plasma Phys.} {\bf 81} 305810104
%
\bibitem{adkins18} Adkins T and Schekochihin AA 2018 {\em J. Plasma Phys.} {\bf 84} 905840107
%
\bibitem{valentini05} Valentini F, Carbone V, Veltri P ... 2005 {\em Phys. Rev. E} {\bf 71} 017402
%
\bibitem{camporeale06} Camporeale E, Delzanno GL, Lapenta G and Daughton W 2006 {\em Phys. Plasmas} {\bf 13} 092110
%
\bibitem{valentini12} Valentini F, Perrone D, Califano F ... 2012 {\em Phys. Plasmas} {\bf 19} 092103
%
\bibitem{pezzi14} Pezzi O, Valentini F and Veltri P 2014 {\em Eur. Phys. J D} {\bf 68} 128
%
\bibitem{camporeale16} Camporeale E, Delzanno GL, Bergen BK and Moulton JD 2016 {\em Comp. Phys. Comm.} {\bf 198} 47-58
%
\bibitem{pezzi16b} Pezzi O, Camporeale E and Valentini F 2016 {\em Phys. Plasmas} {\bf 23} 022103
%
\bibitem{vencels16} Vencels J, Delzanno GL, Manzini G, Markidis S, Peng IB and Roytershteyn V 2016 {\em J. Phys. Conf. Series} vol. 719, 
No. 1, p. 012022 https://doi.org/10.1088/1742-6596/719/1/012022
%
\bibitem{hammett93} Hammett GW, Beer MA, Dorland W, Cowley SC and Smith SA 1993 {\em Plasma Phys. Control. Fusion} {\bf 35} 973-985
%
\bibitem{schekochihin08} Schekochihin AA, Cowley SC, Dorland W, Hammett GW, Howes GG, Plunk GG, Quataert E and Tatsuno T 2008 {\em Plasma 
Phys. Control. Fusion} {\bf 50} 124024
%
\bibitem{tatsuno09} Tatsuno T, Dorland W, Schekochihin AA, Plunk GG, Barnes M, Cowley SC and Howes GG 2009 {\em Phys. Rev. Lett.} {\bf 103} 
015003
%
\bibitem{zocco11} Zocco A and Schekochihin AA 2011 {\em Phys. Plasmas} {\bf 18} 102309
%
\bibitem{teaca12} Teaca B, Navarro AB, Jenko F, Brunner S and Villard L 2012 {\em Phys. Rev. Lett.} {\bf 109} 235003
%
\bibitem{hatch13} Hatch DR, Jenko F, Navarro AB and Bratanov V 2013 {\em Phys. Rev. Lett.} {\bf 111} 175001
%
\bibitem{parker16} Parker JT, Highcock EG, Schekochihin AA and Dellar PJ 2016 {\em Phys. Plasmas} {\bf 23} 070703
%
\bibitem{schekochihin16} Schekochihin AA, Parker JT, Highcock EG, Dellar PJ, Dorland W and Hammett GW 2016 {\em J. Plasma Phys.} {\bf 82} 
905820212
%
\bibitem{watanabe02} Watanabe TH and Sugama H 2002 {\em Phys. Plamsas} {\bf 9} 3659
%
\bibitem{watanabe04} Watanabe TH and Sugama H 2004 {\em Phys. Plasmas} {\bf 11} 1476
%
\bibitem{eyink18} Eyink GL 2018 {\em Cascades and Dissipative Anomalies in Nearly Collisionless Plasma Turbulence} eprint arXiv:1803.03691, http://adsabs.harvard.edu/abs/2018arXiv180303691E
%
\bibitem{landau36} Landau LD 1936 {\em Phys. Z. Sovjet} {\bf 154}, translated in {\em Collected papers of L.D. Landau} edited by D. ter 
Haar pp 163--170 (Oxford: Pergamon, 1981).
%
\bibitem{hirvijoki16} Hirvijoki E, Lingam M, Pfefferl\'e, Comisso L, Candy J and Bhattacharjee A 2016 {\em Phys. Plasmas} {\bf 23} 080701
%
\bibitem{hirvijoki17} Hirvijoki E, Brizard AJ and Pfefferl\'e D 2017 {\em J. Plasma Phys.} {\bf 83} 595830102 
%
\bibitem{pfefferle17} Pfefferl\'e D, Hirvijoki E and Lingam M 2017 {\em Phys. Plasmas} {\bf 24} 042118
%
\bibitem{delzanno15} Delzanno GL 2015 {\em J. Plasma Phys.} {\bf 301} 338--356
%
\bibitem{servidio17} Servidio S, Chasapis A, Matthaeus WH, Perrone D, Valentini F, Parashar TN, Veltri P, Gershman D, Russell CT,Giles B, 
Fuselier SA, Phan TD and Burch J 2017 {\em Phys. Rev. Lett.} {\bf 119} 205101
%
\bibitem{pezzi18} Pezzi O, Servidio S, Perrone D, Valentini F, Sorriso-Valvo L, Greco A, Matthaeus W and Veltri P 2018 {\em Phys. Plasmas} 
{\bf 25} 060704
%
\bibitem{cerri18} Cerri SS, Kunz MW and Califano F 2018 {\em Astrophys. J. Lett.} {\bf 856} L13
%
\bibitem{perrone13} Perrone D, Valentini F, Servidio S, Dalena S and Veltri P 2013 {\em Astrophys. J.} {\bf 762} 99
%
\bibitem{matthaeus14} Matthaeus WH, Oughton S, Osman KT, Servidio S, Wan M, Gary SP, Shay MA, Valentini F, Roytershteyn V and Karimabadi H 2014 {\em Astrophys. J.} {\bf 790} 1557
%
\bibitem{servidio14} Servidio S, Osman KT, Valentini F, Perrone D, Califano F, Chapman S, Matthaeus WH and Veltri P 2014 {\em Astrophys. J. 
Letters} {\bf 781} L27
%
\bibitem{valentini16} Valentini F, Perrone D, Stabile S, Pezzi O, Servidio S, De Marco R ... and Veltri P 2016 {\em New J. Physics} {\bf 
18} 125001
%
\bibitem{valentini17} Valentini F, Vasconez CL, Pezzi O, Servidio S, Malara F and Pucci F 2017 {\em Astron. and Astrophys.} {\bf 599} A8
%
\bibitem{cerri17} Cerri SS, Servidio S and Califano F 2017 {\em Astrophys. J. Letters} {\bf 846} L18
%
\bibitem{groselj17} Groselj D, Cerri SS, Navarro AB, Willmott C, Told D, Loureiro NL, Califano F and Jenko F 2017 {\em Astrophys. J.} {\bf 
847} 28
%
\bibitem{pezzi17b} Pezzi O, Malara F, Servidio S, Valentini F, Parashar TN, Matthaeus WH and Veltri P 2017 {\em Phys. Rev E} {\bf 96} 023201
%
\bibitem{pucci17} Pucci F, Servidio S, Sorriso-Valvo L, Olshevsky V, Matthaeus WH, Malara F, Goldman MV, Newman DL and Lapenta G 2017  {\em 
Astrophys. J.} {\bf 841} 60
%
\bibitem{perrone18} Perrone D, Passot T, Laveder D, Valentini F, Sulem PL, Zouganelis I, Veltri P and Servidio S 2018 {\em Phys Plasmas} 
{\bf 25} 052302
%
\bibitem{parashar18} Parashar TN, Matthaeus WH and Shay MA 2018 {\em Astrophys. J. Letters} {\bf 864} L21
%
\bibitem{franci18} Franci L, Landi S, Verdini A, Matteini L and Hellinger P 2018 {\em Astrophys. J.} {\bf 853} 26
%
\bibitem{hellinger18} Hellinger P, Verdini A, Landi S, Franci L and Matteini L 2018 {\em Astrophys. J. Letters} {\bf 857} L19
%
\bibitem{olshevsky18} Olshevsky V, Servidio S, Pucci F, Primavera L and Lapenta G 2018 {\em Astrophys. J.} {\bf 860} 11
%
\bibitem{pucci18} Pucci F, Matthaeus WH, Chasapis A, Servidio S, Sorriso-Valvo L, Olshevsky V, Newman DL, Goldman MV and Lapenta G 2018 
{\em Astrophys. J.} {\bf 867} 10
%
\bibitem{burch16} Burch JL, Moore TE, Torbert RB and Giles BL 2016 {\em Space Sci. Rev.} {\bf 199} 5
%
\bibitem{rosenbluth57} Rosenbluth MN, MacDonald WM and Judd DL 1957 {\em Phys. Review} {\bf 107} 1
%
\bibitem{dougherty64} Dougherty JP 1964 {\em Phys. Fluids} {\bf 7} 1788 
%
\bibitem{dougherty67a} Dougherty JP and Watson SR 1967 {\em J. Plasma Phys.} {\bf 1} 317--326 
%
\bibitem{dougherty67b} Dougherty JP, Watson SR and Hellberg MA 1967 {\em J. Plasma Phys.} {\bf 1}, 327--339
%
\bibitem{gradshteyn} Gradshteyn IS and Ryzhik IM 2014 {\em Table of integrals, series, and products} (Academic press, New York)
%
\bibitem{pezzi15a} Pezzi O, Valentini F and Veltri P 2015 {\em J. Plasma Phys.} {\bf 81(1)} 305810107
%
\bibitem{pezzi15b} Pezzi O, Valentini F and Veltri P 2015 {\em Phys. Plasmas} {\bf 22(4)} 042112
%
\bibitem{pezzi18b} Pezzi O et al, {\it Collisional effects in solar wind turbulent plasmas: hybrid Vlasov-Maxwell-Dougherty simulations} 
(in preparation).
%
\bibitem{matthaeus82} Matthaeus WH and Goldstein ML 1982 {\em J. Geophys. Res.} {\bf 87(A8)} 6011--6028
%
\bibitem{sorriso99} Sorriso-Valvo L, Carbone V and Veltri P 1999 {\em Geophys. Res. Lett.} {\bf 26(13)} 1801--1804
%
\bibitem{matthaeus05} Matthaeus WH, Dasso S, Weygand JM, Milano LJ, Smith CW and Kivelson MG 2005 {\em Phys. Rev. Lett.} {\bf 95} 231101
%
\bibitem{lagage83} Lagage PO and Cesarsky CJ 1983 {\em Astron. Astrophys.} {\bf 125} 249--257
%
\bibitem{falchetto04} Falchetto GL and Ottaviani M 2004 {\em Phys. Rev. Lett.} {\bf 92} 025002
%
\bibitem{falchetto08} Falchetto GL, Scott BD , Angelino P, Bottino A, Dannert T, Grandgirard V, ... and Romanelli M 2008 {\em Plasma Phys. 
Control. Fusion} {\bf 50} 124015

\end{thebibliography}
\end{document}